\documentclass[10pt]{article}

\newenvironment {acknowledgement}
  {\section*{Acknowledgements}}
  {}
  
\def\sign{{\rm sign}}

\input psfig.sty

\begin{document}

\title
{Generalized thermostatistics and Kolmogorov-Nagumo averages}

\author{
Jan Naudts\footnote{Departement Natuurkunde, Universiteit Antwerpen UIA,
Universiteitsplein 1, B2610 Antwerpen, Belgium,
E-mail: Jan.Naudts@ua.ac.be}
\ and
Marek Czachor\footnote{Katedra Fizyki Teoretycznej i Metod Matematycznych,
Politechnika Gda\'{n}ska, 80-952 Gda\'{n}sk, Poland,
E-mail: mczachor@pg.gda.pl}
}
\date{v8, September 2001}
\maketitle

\begin{abstract}

We introduce a generalized thermostatistics based on
Kol\-mogor\-ov-Nag\-umo averages and appropriately selected 
information measures. The formalism includes Tsallis
non-extensive thermostatistics, but also extensive thermostatistics 
based on R\'enyi entropy. The Curie-Weiss model is discussed
as an example.

Keywords: generalized thermostatistics, Kolmogorov-Nagumo mean, R\'enyi 
entropy, Tsallis entropy, kappa distribution, 
nonlinear averages, non-extensive thermodynamics, mean-field model.

\end{abstract}

\section{Introduction}

We present a thermostatistical theory based on the notion of 
Kolmogorov-Nagumo (KN) mean \cite {RA76}. It deals with the 
problem of maximizing average information content under the 
constraint that the (nonlinear) average of some energy function
has a given value. 

From the point of view of information theory and the principle 
of maximal entropy \cite {JE57} there are at least two ways to 
generalize the Boltzmann-Gibbs formalism of statistical 
mechanics. One can modify the definition of entropy/information, 
or one can modify the definition of average energy. We will make 
use of both possibilities. Many definitions of entropy can be 
found in literature \cite {WA78}. In contrast, KN-averages, 
although known in statistics since many years, appear not to be 
used in the context of physics. Therefore, we will discuss in 
detail how they fit with basic principles of statistical 
mechanics.

Originally, our plan was to generalize the Boltzmann-Gibbs 
formalism by replacing Shannon's entropy by R\'enyi's
$\alpha$-entropies, and by simultaneously replacing the usual linear 
statistical average by an appropriate KN-average.
The resulting formalism is
what we call R\'enyi thermostatistics in the sequel.
However, the formalism allows further generalization
which consists of replacing R\'enyi entropies by equivalent entropies,
each time choosing appropriate KN-averages. By equivalent entropies we
mean monotonically increasing continuous functions of R\'enyi entropies.
Indeed, replacing an entropy by an equivalent one does not
change the maximum entropy problem. One such entropy which is equivalent
with R\'enyi's, and which is quoted often in literature, is
Tsallis' entropy \cite{TC88}, studied earlier by Harvda
and Charvat \cite{HC67} and by Dar\'oczy \cite{DZ70}.
The thermostatistics which we propose is not equivalent
with the non-extensive thermostatistics proposed by
the Tsallis school because of the non-linear averages
used in the present paper.
R\'enyi thermostatistics is extensive, as we will show.
The more general formalism can be non-extensive,
and, in fact, comprises Tsallis' thermostatistics as
a special case.

Note that throughout the paper units are used in which 
Boltzmann's constant $k_B$ equals one.

The paper is organized as follows. In section 2 the general 
formalism is explained. The standard Boltzmann-Gibbs formalism 
and some other formalisms are shown to be special cases. In 
section 3 equilibrium distributions are discussed. Special 
attention is paid to the definition of thermodynamic 
temperature. Section 4 shows how nonlinear averages fit with the 
fundaments of statistical mechanics. The Curie-Weiss model is 
treated as an example. Section 5 deals with R\'enyi 
thermostatistics and its obvious non-extensive generalization. 
The Curie-Weiss model serves again as an example, together with 
the two-level system. The last section contains a short 
discussion of our results. Throughout the paper we use deformed 
exponential and logarithmic functions with a definition which is 
broader than that found in literature. Details about these are 
found in appendix.

\section{A general formalism}

\subsection{Optimization problem}

For simplicity we restrict ourselves initially to functions $f(k)$
with integer $k=0,1,2,\cdots$. We use the index notation $f_k$.

The Kolmogorov-Nagumo (KN) average \cite{KA30, NM30} depends
on a monotonically increasing function $\phi(x)$ and on parameters $p_k$.
Essential for the present approach
is that the latter are probabilities satisfying $p_k\ge 0$ and $\sum_kp_k=1$.
The definition is
\begin{equation}
\langle f\rangle=\phi^{-1}\left(
\sum_kp_k\phi(f_k).
\right)
\label{KN}
\end{equation}
We will not make use of the possibility that $\phi$ may be decreasing
instead of increasing.

Fix an additional monotonically increasing function $\omega(x)$.
It is used to
define information content $I_k$ by
\begin{equation}
I_k=\omega(1/p_k).
\end{equation}
Because $\omega$ is increasing, less probable events
(i.e.~smaller $p_k$) carry more information. The average
information content is now given by
\begin{equation}
\langle I\rangle=\phi^{-1}\left(
\sum_kp_k\phi(\omega(1/p_k))\right).
\label{avinfcont}
\end{equation}

Generalized thermostatistics studies the problem of maximizing
average information content under the condition that average values of functions
$f_1(x),\cdots f_n(x)$ have predetermined values $y_1,\cdots,y_n$
\begin{equation}
\langle f_j\rangle=y_j
\qquad\hbox{ for }j=1,2,\cdots,n.
\end{equation}

Throughout the paper we consider the optimization problem with exactly 
one constraint, i.e., we fix some real function $\beta_0E_k$ and
optimize information content under the condition that
$\langle \beta_0E\rangle$ has a predetermined value $\beta_0U$.
It is tradition to call $E_k$ the energy 
functional and to call $U$ the internal energy. Note that we introduce 
here a constant $\beta_0$ with dimension of inverse energy. Its first 
purpose is to make quantities dimensionless. Its actual role in the 
theory will be clarified later on.
The maximum attained by $\langle I\rangle$
is denoted $S(U)$ and is called thermodynamic entropy
(not to be confused with the entropy functionals introduced below).
The variation is with respect to the choice of probabilities $p_k$.
In principle, it might be interesting to vary also the choice
of functions $\phi(x)$ resp.~$\omega(x)$. This extension is not
considered in the present paper.

\subsection{Variation principle}

It is tradition \cite {JE57} to solve this kind of
optimization problem by introduction
of two Lagrange parameters $\gamma$ and $\beta$, one to ensure
normalization of the probabilities $p_k$, the other to enforce the
energy constraint. Since $\phi$ is a monotonically increasing function
an equivalent optimization problem is to minimize a free
energy of the form
\begin{equation}
\phi(\langle \beta_0 E\rangle)
-\frac{\beta_0}{\beta}\,\phi(\langle I\rangle)
-\gamma\sum_kp_k.
\end{equation}
The set of conditions for an extremum reads
\begin{equation}
0=\frac{\partial\,}{\partial p_k}\phi(\langle \beta_0E\rangle)
-\frac{\beta_0}{\beta}\,\frac{\partial\,}{\partial p_k}\phi(\langle I\rangle)
-\gamma.
\end{equation}
Using (\ref{KN}) and (\ref{avinfcont}) this simplifies to
\begin{equation}
\frac{\partial\,}{\partial p_k}
\left[
p_k\phi(\omega(1/p_k))
\right]
=\frac{\beta}{\beta_0}\left(\phi(\beta_0E_k)-\gamma\right).
\label{varprin}
\end{equation}
This equation has to be solved to obtain $p_k$ as
a function of $\gamma$ and $\beta$.
It is not easy to progress with a problem of this generality.
Let us therefore consider first some important examples.

\subsection{Boltzmann's entropy}

Consider first the rather trivial case that the energy function $E_k$ is
constant and that the index $k$ takes on a finite number $\Omega$
of values. Then entropy is optimal with
\begin{equation}
p_k=\frac{1}{\Omega}.
\end{equation}
This implies immediately the result
\begin{equation}
S=\omega(\Omega).
\end{equation}
With $\omega(x)$ replaced by $k_B\ln(x)$ this is the famous
result obtained by Boltzmann in the nineteenth century.

\subsection{Shannon's entropy}

Let $\phi(x)=x$ and $\omega(x)=\log_b(x)$.
With this choice of $\omega$ the information measure is
\begin{equation}
I_k=-\log_b p_k.
\label{hartley}
\end{equation}
It represents Hartley's notion of information \cite{HRV28}.
The expression for average information content becomes
\begin{equation}
\langle I\rangle=-\sum_kp_k\log_bp_k\equiv S^{\rm Shannon}(p).
\label{shannonentr}
\end{equation}
This expression is called Shannon's entropy functional \cite {SCE48}.
Eq.~(\ref{varprin}) becomes
\begin{equation}
-1-\log_bp_k=\beta E_k-\frac{\beta}{\beta_0}\gamma.
\end{equation}
This implies
\begin{equation}
p_k=\frac{1}{Z}\,\exp_b(-\beta E_k)
\label{gibbs}
\end{equation}
with normalisation factor $Z$ given by
\begin{equation}
Z=\sum_k\exp_b(-\beta E_k).
\end{equation}
This is the Boltzmann-Gibbs distribution. For
simplicity, only the base $b=e$ will be used in the sequel.
The notations $\ln_\alpha(\cdot)$ and $\exp_\alpha(\cdot)$
will be used from now on with the different meaning of $\alpha$-deformed
logarithm resp.~exponential.

\subsection{Tsallis' entropy functional}

Fix some $\alpha>0$, $\alpha\not=1$,
and let $\phi(x)=x$ and $\omega(x)=\ln_\alpha(x)$
where
\begin{equation}
\ln_\alpha(x)=\frac{x^{1-\alpha}-1}{1-\alpha},
\qquad x\ge 1.
\label{defln}
\end{equation}
We use here the $\alpha$-deformed logarithm (\cite {TC94,BEP98},
see also the appendix).
In the limit $\alpha=1$ it reduces to the natural logarithm.
Its inverse is the deformed exponential
\begin{equation}
\exp_\alpha(x)=[1+(1-\alpha)x]_+^{1/(1-\alpha)},
\qquad x\ge 0.
\end{equation}
Throughout the text $[x]_+$ equals the maximum of $x$ and 0.
Note that the present example is an obvious generalization of
the previous one. The average information content is
\begin{equation}
\langle I\rangle
=\sum_kp_k\ln_\alpha(1/p_k)
=\frac{\sum_k p_k^\alpha-1}{1-\alpha}\equiv S_\alpha^{\rm Tsallis}(p).
\label{tsallisentropy}
\end{equation}
This is Tsallis' entropy \cite{HC67,DZ70,TC88}.
Eq.~(\ref{varprin}) becomes
\begin{equation}
\frac{1}{1-\alpha}\left(
\alpha p_k^{\alpha-1}-1
\right)
=\beta E_k-\frac{\beta}{\beta_0}\gamma.
\label{tsallistemp}
\end{equation}
It can be written as
\begin{equation}
p_k=\left[\frac{1}{\alpha}\left(
1+(1-\alpha)\beta (E_k-\gamma/\beta_0)
\right)\right]^{1/(\alpha-1)}.
\end{equation}
which means that $p_k$ is inversely proportional to
$\exp_\alpha(\beta E_k)$.
This solution, with some modification,
will be discussed in more detail further on.

\subsection{R\'enyi's entropy}

It is well-known \cite{RA76_1} that averaging information $I$,
as given by (\ref{hartley}),
leads to R\'enyi's entropy of order $\alpha$, if the function
$\phi$ is chosen of the form
$\phi(x)=\exp((1-\alpha)x)$, ($\alpha\not=1$).
Let us make an innocent modification to this
function and let $\phi=\phi_\alpha$ with
\begin{eqnarray}
\phi_\alpha(x)&=&\ln_\alpha(e^x)\cr
&=&\frac{e^{(1-\alpha)x}-1}{1-\alpha},
\qquad x>0.
\label{phirenyi}
\end{eqnarray}
Note that the KN-mean is invariant under this kind of modification.
One obtains (assuming $\omega(x)=\ln(x)$ as in the Shannon case)
\begin{eqnarray}
\langle I\rangle
&=&\phi^{-1}_\alpha\left(\sum_kp_k\phi_\alpha(-\ln p_k)\right)\cr
&=&\phi^{-1}_\alpha\left(\frac{\sum_kp_k^\alpha-1}{1-\alpha}\right)\cr
&=&\frac{1}{1-\alpha}\ln\left(\sum_kp_k^\alpha\right)
\equiv S_\alpha^{\rm Renyi}(p)
\label{renyi}
\end{eqnarray}
The latter is the usual expression for R\'enyi's entropy of order $\alpha$.

It is now straightforward to solve (\ref{varprin}) for the present 
choice of $\phi$ and $\omega$. However, note that in course of calculation 
(\ref{renyi}) the expression for Tsallis' entropy (\ref{tsallisentropy}) 
appears. This shows that in the context of KN-means, there is an intrinsic
relation between the entropies of R\'enyi and of Tsallis
(this relation seems to be known since quite some time --- see \cite {VP01}).
Instead of maximizing R\'enyi's entropy under the constraint that
$\langle \beta_0E\rangle=\beta_0U$ one can as well maximize
Tsallis' entropy $\phi_\alpha(\langle I\rangle)$
under the constraint that $\phi_\alpha(\langle \beta_0E\rangle)
=\phi_\alpha(\beta_0U)$. In other words, the
present example is reduced to the previous one.

\subsection{Equivalent entropies}

Note that there exist other pairs $(\phi,\omega)$ which are linked
to Tsallis entropy in a way similar to the previous example.
This is the case whenever
\begin{equation}
\phi(\omega(x))=\ln_\alpha(x)
\label{assume}
\end{equation}
for some $\alpha\not=1$.
One can see (\ref{assume}) as a definition of $\omega$
in case $\phi$ is given, or as a constraint on $\phi$
if $\omega$ is given.

Condition (\ref{assume}) implies that $\phi$ maps average information 
content $\langle I\rangle$ onto Tsallis' entropy.
Hence all entropies in this class are equivalent in the sense
that they are increasing functions
of Tsallis' $\alpha$-entropy or, if you wish, of R\'enyi's $\alpha$-entropy.
Note that the result of the optimization problem does not change
if the entropy is replaced by a monotonic function of the entropy.
Hence all what happens is that for a given (nonlinear)
constraint we adapt the $\alpha$-entropy to an equivalent entropy which
is technically better adapted for solving the optimization problem.

\section{Solutions}

In this section $\alpha$ is fixed and $\phi$ and $\omega$ are related 
by (\ref{assume}). 

The original optimization problem is reformulated as
maximizing Tsallis' entropy $S_\alpha^{\rm Tsallis}(p)$
under the constraint
\begin{equation}
\sum_kp_k \phi(\beta_0E_k)=\phi(\beta_0 U)
\label{constraint}
\end{equation}

\subsection{Optimization results}

The formulas that follow are based on results already found
in literature in many places, e.g.~in \cite{NJ00}.

The equilibrium average
is the KN-mean (\ref{KN}) with $p_k$ given by
\begin{equation}
p_k=\frac{1}{Z_1}\,[1+a(1-\alpha)(x_k-u)]_+^{1/(\alpha-1)}
=\frac{1}{Z_1}\,\exp_\alpha(-a(x_k-u))
\label{alphaless}
\end{equation}
with $a>0$ and $u\le x_k$ for all $k$, with
normalization $Z_1$ given by
\begin{equation}
Z_1=\sum_k[1+a(1-\alpha)(x_k-u)]_+^{1/(\alpha-1)}
\label {Z1def}
\end{equation}
and with $x_k=\phi(\beta_0 E_k)$.
The unknown parameters $a$ and $u$ have to
be fixed in such a way that (\ref{constraint}) holds.
This condition can be written as
\begin{eqnarray}
\phi(\beta_0 U)&=&u+\frac{1}{a(1-\alpha)}\left(\frac{Z_0}{Z_1}
-1\right)
\label{uless}
\end{eqnarray}
with $Z_0$ given by
\begin{equation}
Z_0=\sum_k[1+a(1-\alpha)(x_k-u)]_+^{\alpha/(\alpha-1)}.
\label{z0less}
\end{equation}
There is only one condition to determine the two parameters
$a$ and $u$. A convenient way to fix $u$ is to take it equal to
the lower bound of the $x_k$ (we assume that $x_k$ is bounded from below).
In that case one can guarantee \cite{NJ00} that the solution of
the optimization problem, if it exists, occurs for $a\ge 0$.

The entropy $S(U)=\langle I\rangle_{\rm max}$
follows from (\ref{tsallisentropy})
using (\ref{alphaless}). One obtains
\begin{equation}
\phi(S)=\frac{1}{1-\alpha}\left(\frac{Z_0}{Z_1^\alpha}-1\right).
\label{sless}
\end{equation}

Expression (\ref{alphaless}) with $0<\alpha<1$ is related to
the kappa-distribution or generalized Lorentzian
distribution \cite{MZ00}
\begin{equation}
p(v)=\frac{1}{\left[1+av^2\right]^{1+\kappa}}.
\end{equation}
It is frequently used, e.g.~in plasma physics to describe an excess
of highly energetic particles \cite{MVMH95}.

Typical for distribution (\ref{alphaless}) with $\alpha>1$
is that the probabilities $p_k$ 
are identically zero whenever $a(\alpha-1)(\phi(E_k)-u)\ge 1$.
This cut-off for high values of $E_k$ is of interest in many areas of physics.
In astrophysics it has been used \cite{PP93} to describe stellar systems
with finite average mass.
A statistical description of an electron captured
in a Coulomb potential requires the cut-off to mask scattering
states \cite{LST95,NJ01}. In standard statistical mechanics the treatment of
vanishing probabilities requires infinite energies which lead to
ambiguities. These can be avoided if distributions of the type (\ref{alphaless}),
with $\alpha>1$, are used.

\subsection{A thermodynamic relation}

For further use we calculate now the derivative of $S$
w.r.t~$U$. Since the equilibrium state depends only on
the fee parameter $a$ ($u$ may be kept constant) we
start with calculating the dependence of $S$ on $a$.

Equations (\ref{uless}) and (\ref{sless}) can be written as
\begin{eqnarray}
\frac{Z_0}{Z_1^\alpha}&=&1+(1-\alpha)\phi(S)\cr
\frac{Z_0}{Z_1}&=&1+a(1-\alpha)(\phi(\beta_0 U)-u)
\end{eqnarray}
Hence, (\ref{sless}) can be written as
\begin{equation}
\phi(S)=\frac{1}{1-\alpha}\left(
[1+a(1-\alpha)(\phi(\beta_0U)-u)]^\alpha Z_0^{1-\alpha}-1
\right)
\end{equation}
On the other hand, (\ref{z0less}) implies
\begin{equation}
\frac{{\rm d}\,}{{\rm d}a}Z_0=-\frac{\alpha}{1-\alpha}\frac{1}{a}(Z_0-Z_1)
\label{tempdera}
\end{equation}
Combining these two expressions one obtains
\begin{eqnarray}
\frac{{\rm d}\,}{{\rm d}a}\phi(S)
&=&a\alpha
\frac{1+(1-\alpha)\phi(S)}{1+a(1-\alpha)(\phi(\beta_0 U)-u)}
\frac{{\rm d}\,}{{\rm d}a}\phi(\beta_0 U)
\end{eqnarray}
The final result is then 
\begin{eqnarray}
\frac{1}{\beta_0}\,
\frac{{\rm d}S}{{\rm d}U}
&=&\frac{{\rm d}S}{{\rm d}a}\bigg/ \beta_0\frac{{\rm d}U}{{\rm d}a}\cr
&=&\frac{\phi'(\beta_0U)}{\phi'(S)}\left(
\frac{{\rm d}\phi(S)}{{\rm d}a}\bigg/ \frac{{\rm d}\phi(\beta_0U)}{{\rm d}a}
\right)\cr
&=&a\alpha \,\frac{\phi'(\beta_0U)}{\phi'(S)}\,
\frac{1+(1-\alpha)\phi(S)}{1+a(1-\alpha)(\phi(\beta_0 U)-u)}
\label{thermorel}
\end{eqnarray}
Note that $a$ in this expression is still a function
of $U$.

\subsection{Thermodynamic temperature}

The standard thermodynamical relation for temperature $T$ is
\begin{equation}
\frac{1}{T}=\frac{{\rm d}S}{{\rm d}U}.
\label{standtemp}
\end{equation}
In generalized thermostatistics
this definition of temperature is not necessarily correct
\cite {AMPP01}. The problem is
that the thermodynamic definition of temperature
(\ref{standtemp}) is not invariant under substitution
of entropy by an equivalent entropy. This means that one can
generalize definition (\ref{standtemp}) to
\begin{eqnarray}
\frac{1}{T}
&=&\frac{{\rm d}f(S)}{{\rm d}U}
=f'(S)\frac{{\rm d}S}{{\rm d}U}
\label{gendeftemp}
\end{eqnarray}
where $f(x)$ is any strictly increasing function of $x$
i.e.~the derivative $f'(x)$ is positive. However, not all
choices of $f(x)$ are physically acceptable.
Minimal requirements are that $T$ is an increasing
function of $U$, and that $T=0$ corresponds with
$U=E_{\rm m}$, where $E_{\rm m}=\min_kE_k$.

The formalism of Tsallis' thermostatistics\cite {TC88}
has been modified a few times \cite{CT91,TMP98}.
In particular, the correct definition of temperature has
been a difficult point \cite{AMPP01,TR01,VP01}. 
One can easily verify that the proposed definitions
of thermodynamic temperature are of the form (\ref{gendeftemp}).
As argued in \cite{AMPP01},
(\ref{standtemp}) is correct in case the entropy is extensive. Only
R\'enyi's $\alpha$-entropies are extensive -- see the next
section. Hence we propose that $f(x)$ should be fixed in
such a way that $f(S)$ is R\'enyi's entropy. This is the case
if
\begin{equation}
f(x)=\ln(\exp_\alpha(\phi(x)))=\frac{1}{1-\alpha}\ln(1+(1-\alpha)\phi(x))
\label{fdef}
\end{equation}
Indeed, one has
\begin{equation}
\phi(S)=S_{\rm Tsallis}=\ln_\alpha(\exp(S_{\rm Renyi})).
\end{equation}
Note that, taking $\phi(x)=x$ in (\ref{fdef}), one
obtains
\begin{eqnarray}
\frac{1}{T}
&=&\frac{{\rm d}f(S)}{{\rm d}U}
=\frac{1}{1+(1-\alpha)S}\frac{{\rm d}S}{{\rm d}U}.
\end{eqnarray}
This coincides with the formula
proposed in \cite{AMPP01}.

\subsection{Discussion}

Combine (\ref{thermorel},\ref{gendeftemp}) with the above
choice of function $f(x)$, one obtains
\begin{eqnarray}
\frac{1}{\beta_0T}
=
\frac{a\alpha\phi'(\beta_0U)}{1+a(1-\alpha)(\phi(\beta_0 U)-u)}.
\label{gentempexpr}
\end{eqnarray}
Make now the following choice of the parameter $u$
\begin{equation}
u=\phi(\beta_0E_{\rm m})
\qquad \hbox{ with }E_{\rm m}=\min_k E_k.
\end{equation}
Then (\ref{gentempexpr}) can be written as
\begin{eqnarray}
\frac{1}{\beta_0T}
=
\frac{a\alpha\phi'(\beta_0U)}
{1+a(1-\alpha)(\phi(\beta_0 U)-\phi(\beta_0 E_{\rm m}))}.
\label{gentempexpr2}
\end{eqnarray}
This is our general result for the relation between the
parameter $a$ and thermodynamic temperature $T$.
From this expression it is immediately clear that,
in case $0<\alpha<1$, temperature $T$ is always positive.
The analysis of what happens if $\alpha>1$
is more complex and falls out of the scope of the present paper.
In addition, if $a$ tends to infinity, then $U$
will become equal to $E_{\rm m}$. In this limit (\ref{gentempexpr2})
implies that temperature $T$ goes to zero, as is physically
expected.

An open questions is wether $T$ is always an increasing function of $U$,
as it should be. This analysis is not made in the present paper.

Next, use (\ref{gentempexpr}) to eliminate $a$ from
expression (\ref{alphaless}) for the probabilities $p_k$.
One obtains
\begin{eqnarray}
p_k=\frac{1}{\zeta_\alpha}\left[
\alpha+(1-\alpha)\frac{1}{\phi'(\beta_0U)\beta_0T}(\phi(\beta_0E_k)-\phi(\beta_0U))
\right]_+^{1/(\alpha-1)}
\label{ftdistr}
\end{eqnarray}
with $\zeta_\alpha$ the appropriate normalization constant.
The limit $\alpha=1$ of this expression is
\begin{equation}
p_k=\frac{1}{\zeta_1}\exp\left[1-
\frac{1}{\phi'(\beta_0U)\beta_0T}(\phi(\beta_0E_k)-\phi(\beta_0U))\right].
\label{phidistr}
\end{equation}
If $\phi(x)=x$ then this is the Boltzmann-Gibbs distribution.

\subsection{A special temperature}

Consider the case that
\begin{equation}
a=\frac{1}{1+(1-\alpha)u}
\end{equation}
Then (\ref{alphaless}) simplifies to
\begin{equation}
p_k=\frac{1}{\zeta}\exp_\alpha(-x_k)
\label{genbgdist}
\end{equation}
with $\zeta$ the appropriate normalization factor.
Formula (\ref{gentempexpr2}) becomes
\begin{equation}
\frac{1}{\beta_0T}
=\frac{\alpha\phi'(\beta_0U)}{1+(1-\alpha)\phi(\beta_0U)}
\label{spectemp}
\end{equation}
In case of R\'enyi thermostatistics is $\phi=\phi_\alpha$
as given by (\ref{phirenyi}). Then the r.h.s.~of the above
expression is identically equal to $\alpha$ so that one obtains the
condition $\alpha\beta_0T=1$. This shows that in this case the
special temperature for which (\ref{genbgdist}) holds
is $1/\alpha\beta_0$. One concludes that, because of the
nonlinear averages, the value of $\beta_0$ cannot be chosen
arbitrarily.

\section{Nonlinear averages}

In the present section we try to clarify the role nonlinear
averages can play in statistical mehanics.

\subsection{Large deviations}

A fundament of equilibrium statistical mechanics is the 
equivalence of ensembles, known as the equipartition theorem \cite{GR77}
in information theory. A related \cite{LPS94a,LPS94b,LP95}
property is the asymptotic expression
\begin{equation}
\hbox{Prob}(\beta_0(E-U)/N\approx\epsilon)\sim \exp(-Nr(\epsilon))
\label{largedev}
\end{equation}
where $r(\epsilon)$ is an appropriate rate function.
With '$f_N\sim\exp(-Nx)$' we mean here that
\begin{equation}
\lim_{N\rightarrow\infty}\frac{1}{N}\ln f_N=x,
\end{equation}
with $x\approx\epsilon$ we mean that $x$ should ly in a small open
neighborhood of $\epsilon$, of size independent of $N$.
This result from the theory of large deviations
\cite {ERS85} states that the probability that energy $E$ deviates 
from its average value $U$ is exponentially small in the size of the 
system $N$. Physically, minus the rate function $r(\epsilon)$ has
the meaning of a relative entropy density in the microcanonical ensemble
with energy $U+N\epsilon$.
The rate function $r(\epsilon)$ is expected to be minimal for $\epsilon=0$
because the equilibrium state maximizes entropy under
the constraint that average energy equals $U$.

In case of identically distributed independent stochastic variables
$x_1$, $x_2$, $\cdots$ with average value $\langle x\rangle$,
the central limit theorem implies that
\begin{equation}
\hbox{Prob}\left(\frac{1}{N}\sum_{j=1}^N(x_j-\langle x\rangle)\approx\epsilon
\right)\sim  e^{-N\epsilon^2/2\sigma^2}
\label{clt}
\end{equation}
where $\sigma^2$ is the variance of the distribution.
Comparison with (\ref{largedev}) shows that in this case
$r(\epsilon)=\epsilon^2/2\sigma^2$.
In general, (\ref{largedev}) will hold
with different rate functions. E.g., in the high temperature phase
of the Curie-Weiss model, is $U=\hbox{O}(N^0)$ and
\begin{eqnarray}
\hbox{Prob}(\beta_0E/N\approx\epsilon)
&\sim& \exp\left(-N\left[s_\infty\left(\sqrt{-\frac{2\epsilon}{\beta_0J}}\right)
+\frac{\epsilon}{\beta_0T}\right]
\right),
\label{cwht}
\end{eqnarray}
($-\beta_0J/2<\epsilon<0$),
with $T$ the temperature, with $J>0$ the interaction constant of the model, and with
\begin{eqnarray}
s_\infty(m)
&=&\frac{1}{2}(1+m)\ln(1+m)
+\frac{1}{2}(1-m)\ln(1-m)
\label{infrate}
\end{eqnarray}
($-1<m<1$, see section (\ref{sect_cw}) below).

Let us now make the link with nonlinear averages.
Fix a strictly increasing function $\phi(x)$.
Then (\ref{largedev}) is equivalent with
\begin{equation}
\hbox{Prob}(\phi(\beta_0(E-U)/N)\approx\epsilon)
\sim\exp(-Nr\circ\phi^{-1}(\epsilon))
\label{largedev2}
\end{equation}
For thermostatistics it is important that the rate function $r(\epsilon)$
is convex, at least
in a neighborhood of $\epsilon=0$. But clearly,
that $r(\epsilon)$ is convex does not imply that $r\circ\phi^{-1}(\epsilon)$
is convex. Conversely, if $r(\epsilon)$ is
not convex, as is e.g.~the case in the Curie-Weiss model
at low temperatures,
then one might search for a function $\phi(x)$ such that
$r\circ\phi^{-1}(\epsilon)$ is convex.
This is a possible motivation for introducing KN-averages.
The strictly increasing function $\phi(x)$
changes the scale used to observe energy fluctuations.
The choice of function should be such that the resulting
rate function is convex.

For example, if we take $\phi(x)=\sign(x)\sqrt{|x|}$ and apply rescaling to
(\ref{cwht}) we obtain
\begin{equation}
\hbox{Prob}\left(\sqrt{-\beta_0E/N}\approx\epsilon\right)
\sim\exp(-N\left[s_\infty\left(\epsilon\sqrt{\frac{2}{\beta_0J}}\right)
-\frac{\epsilon}{\beta_0T}
\right]
\label{cwht2}
\end{equation}
Note that in the Curie-Weiss model the square root of energy is 
proportional to total magnetization $M$. In fact, it is 
tradition to study large deviations of the Curie-Weiss model in 
terms of $M$ instead of $E$, in the context of a grand canonical
ensemble instead of the canonical ensemble studied here.

The link with Kolmogorov-Nagumo (KN) averages (see definition (\ref{KN}))
is immediate, be it on an intuitive basis. We introduce probabilities $p_k$ of a 
canonical ensemble based on (\ref{largedev2}), i.e.\ after
rescaling energy fluctuations with the function $\phi$,
because then we can expect equivalence of ensembles to be valid.
The actual need for nonlinear averages, as given by (\ref{KN}),
will be discussed in section \ref{wna_sect}.

\subsection{The Curie-Weis model}
\label{sect_cw}

For a mathematical treatment of the Curie-Weiss model in the 
context of the Boltzmann-Gibbs formalism, see \cite{ERS85, EW90, 
LPS94b}. The energy of the model is given by
\begin{equation}
E_\sigma=-\frac{J}{2N}M_\sigma^2-hM, 
\qquad \hbox{ with } M_\sigma=\sum_{j=1}^N\sigma_j.
\end{equation}
The spin variables $\sigma_j$ take on the values $\pm 1$.
The constant $J$ is strictly positive. For simplicity,
assume $h>0$.

Fix $\gamma>0$. Introduce an increasing function $\phi(x)$ by
\begin{equation}
\phi(x)=\sign(x)|x|^\gamma.
\end{equation}
Because $\phi'(x)$ is singular at $x=0$ when $0<\gamma<1$
we have to be carefull with the denominator $\phi'(U)$
appearing in (\ref{phidistr}).
Let $m=M/N$ and $v=-U/N$ and assume $U<0$. There follows
\begin{eqnarray}
p_\sigma
&=&\frac{1}{\zeta_1}\exp\left[
-\frac{|U|}{\gamma T}
\left(\sign(E_\sigma)\left|
\frac{E_\sigma}{U}\right|^\gamma
+1\right)
\right]\cr
&=&\frac{1}{\zeta_1'}\exp\left[
N\frac{v^{1-\gamma}}{\gamma T}
\phi(Jm^2/2+hm)
\right]
\end{eqnarray}
Hence the probability of a given magnetization $M$ is
\begin{eqnarray}
p(M)
&=&{N\choose \frac{N+M}{2}}\frac{1}{\zeta_1'}\exp\left[
N \frac{v^{1-\gamma}}{\gamma T}
\phi(Jm^2/2+hm)
\right]\cr
&\sim&
\exp(-Nr(m))
\label{apppm}
\end{eqnarray}
with
\begin{eqnarray}
r(m)&=&s_\infty(m)
-\frac{v^{1-\gamma}}{\gamma T}\phi(Jm^2/2+hm)
\label{cwg}
\end{eqnarray}
(see (\ref{infrate}) for the definition of $s_\infty$).
To obtain this asymptotic expression,
Stirling's formula has been used. Note that
(\ref{cwg}) is {\sl not} in agreement with (\ref{cwht},\ref{cwht2})!
The result obtained here is ($\gamma=1/2$)
\begin{equation}
\hbox{Prob}\left(\sqrt{-\beta_0E/N}\approx\epsilon\right)
\sim\exp(-N\left[s_\infty\left(\epsilon\sqrt{\frac{2}{\beta_0J}}\right)
-\frac{\sqrt{2Jv}}{T}\epsilon
\right]
\label{cwht3}
\end{equation}
The temperature $T$ in this expression is defined by
the thermodynamic relation (\ref{standtemp}),
while in (\ref{cwht},\ref{cwht2}) it is a
Lagrange multiplier. Since $v=\hbox{O}(1/N)$ throughout
the high-temperature phase
the second contribution to the rate function in (\ref{cwht3})
vanishes in the limit of large $N$.

Assume $r(m)$ is minimal at $m=m_*$. The condition
$I'(m_*)=0$ reads
\begin{equation}
m_*=\tanh\left(\frac{1}{T}
\left(\frac{v}{Jm_*^2/2+hm_*}
\right)^{1-\gamma}
(Jm_*+h)
\right)
\end{equation}
Because $v=Jm_*^2/2+hm_*+\hbox{O}(1/N)$ holds, this simplifies to
\begin{equation}
m_*=\tanh\left(\frac{1}{T}(Jm_*+h)
\right).
\label{mfeq}
\end{equation}
This is the standard mean-field equation. It does not depend
on the choice of $\gamma$.

Let $m=m_*+\epsilon$. Expansion of the exponent in (\ref{apppm})
around $m_*$, assuming $m_*>0$, gives
\begin{eqnarray}
p(M)\sim\exp[-(1/2)NI''(m_*)\epsilon^2]
\end{eqnarray}
with
\begin{eqnarray}
I''(m_*)&=&\frac{1}{1-m_*^2}-\frac{J}{T}
+(1-\gamma)\frac{1}{T}\frac{(Jm_*+h)^2}{Jm_*^2/2+hm_*}
\label{cwgdd}
\end{eqnarray}
Hence fluctuations do depend on the choice of $\gamma$.

Let us now consider the case $\gamma<1$ and $h=0$ in  more 
detail. Then $m_*=0$ is a solution of the mean-field equation at 
all temperatures. The corresponding value of $v$ is of the order $1/N$
and can be neglected.
Hence expression (\ref{cwg}) reduces to the result of
the ideal paramagnet. One concludes that, contrary to what
happens in the $\gamma=1$-case, this solution
is (meta)stable at all temperatures (a solution of the mean-field
equation is said to be stable if it optimizes entropy for
the given average energy, it is metastable if it is not stable
but the rate function is convex in a neighborhood of the
origin).

\section{R\'enyi thermostatistics}
\label {renyisect}

R\'enyi's entropy corresponds with the choice $\phi=\phi_\alpha$
(given by (\ref{phirenyi}))
and $I$ given by (\ref{hartley}), i.e.~$\phi_\alpha(x)=\ln_\alpha(e^x)$
and $\omega(x)=\ln(x)$. In the present section we study an obvious
generalization of this entropy. It contains both R\'enyi
and Tsallis thermostatistics as subcases.

\subsection {Definition}

Introduce the function $\phi_{\alpha\rho}(x)$ given by
\begin{eqnarray}
\phi_{\alpha\rho}(x)&=&\ln_\alpha(\exp_\rho(x))\cr
&=&\sign(x)\frac{1}{1-\alpha}\left(
[1+(1-\rho)|x|]_+^{\theta(\alpha,\rho)}-1
\right)
\label{alfarodef}
\end{eqnarray}
Here, and in what follows, we use $\theta(\alpha,\rho)$ as an abbreviation for
$(1-\alpha)/(1-\rho)$.
Equation (\ref{alfarodef}) is the obvious generalization of (\ref{phirenyi})
obtained by deforming the exponential occurring in it.
We use extended definitions of $\ln_\alpha(x)$
and $\exp_\rho(x)$, introduced in appendix.
The inverse function of $\phi_{\alpha\rho}$ is $\phi_{\rho\alpha}$.
The entropy function is chosen in such a way that
(\ref{assume}) holds. Now, $\phi_{\alpha\rho}(\omega(x))=\ln_\alpha(x)$
implies $\omega(x)=\ln_\rho(x)$. The average information content
then equals
\begin{eqnarray}
\langle I\rangle
&=&\phi_{\rho\alpha}
\left(\sum_kp_k\phi_{\alpha\rho}(\ln_\rho (1/p_k))\right)\cr
&=&\frac{1}{1-\rho}\left[
\left(\sum_kp_k^\alpha\right)^{\theta(\rho,\alpha)}-1\right]
\equiv S_{\alpha\rho}(p)
\label{twoparam}
\end{eqnarray}
The two interesting limiting cases are $\rho=1$ and $\rho=\alpha$.
If $\rho=1$ the expression reduces to that of R\'enyi's
entropies (\ref{renyi}). 
If $\rho=\alpha$ it reduces to the Tsallis expression (\ref{tsallisentropy}).
Note that all entropies with the same $\alpha$-value belong to the same
equivalence class. The constraint $\langle\beta_0 E\rangle=U$
can be written as
\begin{equation}
\sum_kp_k\phi_{\alpha\rho}(\beta_0E_k)=\phi_{\alpha\rho}(U)
\end{equation}
It depends of course on the choice of both $\alpha$ and $\rho$.

\subsection{Subsystems}

R\'enyi proved \cite{RA76} that the only additive measures of 
information content are Shannon's entropy (\ref{shannonentr}) 
and the family of R\'enyi $\alpha$-entropies (\ref{renyi}). 
Additivity means here that when a system is composed of two 
independent subsystems A and B then the entropy of the total 
system is the sum of the entropies of the subsystems.

Let us consider additivity in more detail. A system
consisting of two independent subsystems is described
by probabilities $p_{jk}$ of the product form
\begin{equation}
p_{jk}=p^{(1)}_jp^{(2)}_k
\end{equation}
A short calculation gives
\begin{eqnarray}
\langle I\rangle^{(1,2)}
&=&\frac{1}{1-\rho}\left[
\left(\sum_{jk}p_{jk}^\alpha\right)^{\theta(\alpha,\rho)}-1
\right]\cr
&=&\frac{1}{1-\rho}\left[
\left(\sum_{j}(p_{j}^{(1)})^\alpha\right)^{\theta(\alpha,\rho)}
\left(\sum_{k}(p_{k}^{(2)})^\alpha\right)^{\theta(\alpha,\rho)}
-1
\right]\cr
&=&\langle I\rangle^{(1)}+\langle I\rangle^{(2)}
+(1-\rho)\langle I\rangle^{(1)} \langle I\rangle^{(2)}
\end{eqnarray}
If $\rho=1$ this implies additivity of R\'enyi's entropy functional.
If $\rho=\alpha$ this result is well-known \cite{TC88} in the literature
about Tsallis' thermostatistics.

Next assume that the relation between the energy functional
of the composed system and those of the subsystems is
\begin{eqnarray}
\beta_0E_{jk}^{(1,2)}
=\ln_\rho\left(\exp_\rho(\beta_0E_j^{(1)})\exp_\rho(\beta_0E_k^{(2)})
\right)
\label{energprod}
\end{eqnarray}
For $\rho=1$ this simply means that $E_{jk}^{(1,2)}=E_j^{(1)}+E_k^{(2)}$.
For arbitrary $\rho$ (\ref{energprod}) implies that
\begin{eqnarray}
\phi_{\alpha\rho}(\beta_0E_{jk}^{(1,2)})
=\ln_{\alpha}\left(
\exp_\rho(\beta_0E_j^{(1)})\exp_\rho(\beta_0E_k^{(2)})
\right)
\end{eqnarray}
so that
\begin{eqnarray}
& &\sum_{jk}p_{jk}\phi_{\alpha\rho}(\beta_0E_{jk})\cr
&=&\frac{1}{1-\alpha}\sum_jp_j^{(1)}\sum_kp_k^{(2)}
\left(\left[
\exp_\rho(\beta_0E_j^{(1)})\exp_\rho(\beta_0E_k^{(2)})
\right]^{1-\alpha}-1
\right)\cr
&=&\frac{1}{1-\alpha}\bigg[
\sum_jp_j^{(1)}\exp_\rho(\beta_0E_j^{(1)})^{1-\alpha}
\sum_kp_k^{(2)}\exp_\rho(\beta_0E_k^{(2)})^{1-\alpha}
-1
\bigg]\cr
&=&\frac{1}{1-\alpha}\bigg[
\sum_jp_j^{(1)}(1+(1-\alpha)\phi_{\alpha\rho}(\beta_0E_j^{(1)})\cr
& &\times
\sum_kp_k^{(2)}(1+(1-\alpha)\phi_{\alpha\rho}(\beta_0E_k^{(2)})
-1\bigg]\cr
&=&X^{(1)}+X^{(2)}+(1-\alpha)X^{(1)}X^{(2)}
\end{eqnarray}
with
\begin{equation}
X^{(i)}=\sum_jp_j^{(i)}\phi_{\alpha\rho}(\beta_0E_j^{(i)})
\end{equation}
On the other hand is
\begin{eqnarray}
& &\langle\beta_0E^{(1)}\rangle^{(1)}
+\langle\beta_0E^{(2)}\rangle^{(2)}
+(1-\rho)
\langle\beta_0E^{(1)}\rangle^{(1)}
\langle\beta_0E^{(2)}\rangle^{(2)}\cr
&=&\ln_\rho(\exp_\alpha(X^{(1)}))
+\ln_\rho(\exp_\alpha(X^{(2)}))\cr
& &
+(1-\rho)
\ln_\rho(\exp_\alpha(X^{(1)}))
\ln_\rho(\exp_\alpha(X^{(2)}))\cr
&=&\ln_\rho
\left\{
\exp_\alpha(X^{(1)})\exp_\alpha(X^{(2)})
\right\}
\cr
&=&\ln_\rho\left(
\left\{
(1+(1-\alpha)X^{(1)})(1+(1-\alpha)X^{(2)})
\right\}^{1/(1-\alpha)}
\right)\cr
&=&\phi_{\rho\alpha}\left(
X^{(1)}+X^{(2)}+(1-\alpha)X^{(1)}X^{(2)}
\right)
\end{eqnarray}
Combination of the two expressions gives the result
\begin{eqnarray}
\langle\beta_0E\rangle
=\langle\beta_0E^{(1)}\rangle^{(1)}
+\langle\beta_0E^{(2)}\rangle^{(2)}
+(1-\rho)
\langle\beta_0E^{(1)}\rangle^{(1)}
\langle\beta_0E^{(2)}\rangle^{(2)}
\end{eqnarray}
For $\rho=1$ this proves the additivity for
energies as expected.
When $\rho\not=1$ this is a nontrivial result,
even if $\rho=\alpha$, in which case the average is linear.

In general, equilibrium probabilities of generalized thermostatistics
are not of the product form. Therefore, thermodynamic
entropy $S$, which is the maximum value attained by $\langle I\rangle$
is not necessarily equal to the sum of thermodynamic entropies of
the subsystems. Of course, the product form is also absent in the standard formalism
of thermostatistics when there are correlations between subsystems. If
correlations are not too strong then the system in equilibrium
is still extensive, by which is meant that thermodynamic entropy $S$
and internal energy $U$ grow linearly with the size of the system.
This is usually expressed by stating that the so-called
thermodynamic limit exists. We expect that also in the present case
(i.e.~with $\phi_\alpha$ given by (\ref{phirenyi}) 
and $I$ given by (\ref{hartley}))
thermodynamic limit exists under suitable conditions.
This point has still to be studied.

\subsection{Two-level system}
\label{2lev_sect}

As an example, consider a system with two energy levels
$E_0$ and $E_1$.
Assume $E_0=0$ and $E_1>0$. Let $u=0$.
Then one has
\begin{eqnarray}
x_0&=&\phi_{\alpha\rho}(\beta_0E_0)=\phi_{\alpha\rho}(0)=0
\qquad\hbox{ and }\cr
x_1&=&\phi_{\alpha\rho}(\beta_0E_1)=\frac{1}{1-\alpha}\left(
[1+(1-\rho)\beta_0E_1]_+^{\theta(\alpha,\rho)}-1
\right)
\label{2levx}
\end{eqnarray}
From (\ref{alphaless}) follows $p_0=1/Z_1$ and
\begin{equation}
p_1=p_0\left[1+a(1-\alpha)x_1\right]_+^{-1/(1-\alpha)}
\label{2levp}
\end{equation}
Note for further use that
\begin{equation}
Z_1=\frac{1}{p_0}=1+\frac{p_1}{p_0}
\qquad\hbox{ and }
Z_0=1+(Z_1-1)^\alpha=1+\left(\frac{p_1}{p_0}\right)^\alpha
\label{exz0z1}
\end{equation}
From
\begin{eqnarray}
\phi_{\alpha,\rho}(\beta_0U)
&=&p_0\phi_\alpha(0)+p_1\phi_\alpha(1)
=p_1x_1
\label{exphien}
\end{eqnarray}
follows
\begin{eqnarray}
1-p_0=p_1
&=&\frac{1}{x_1}\,\phi_{\alpha,\rho}(\beta_0U)\cr
&=&\frac{1}{1+(1-\alpha)x_1}
\left([1+(1-\rho)\beta_0U]_+^{\theta(\alpha,\rho)}-1\right).
\label{exless}
\end{eqnarray}
This is the solution of the optimization problem
maximizing information content given the constraint that
$\langle \beta_0E\rangle=\beta_0 U$.

Entropy can be calculated using (\ref{exz0z1}). One obtains
\begin{eqnarray}
\phi_{\alpha\rho}(S)
&=&\frac{1}{1-\alpha}\left(\frac{Z_0}{Z_1^\alpha}-1\right)\cr
&=&\frac{1}{1-\alpha}\left(p_0^\alpha+p_1^\alpha-1\right)
\end{eqnarray}
so that
\begin{eqnarray}
S&=&\frac{1}{1-\rho}\left(
(p_0^\alpha+p_1^\alpha)^{\theta(\rho,\alpha)}-1
\right).
\end{eqnarray}

The optimum as a function of temperature is calculated now.
From definition (\ref{gendeftemp}) of thermodynamic temperature,
with $f(x)$ given by (\ref{fdef}), there follows
\begin{eqnarray}
\frac{1}{\beta_0T}
&=&\frac{\phi'_{\alpha\rho}(S)}{1+(1-\alpha)\phi_{\alpha\rho}(S)}
\,\frac{{\rm d}S}{{\rm d}\beta_0U}\cr
&=&\frac{1}{1+(1-\rho)S}\,\frac{{\rm d}S}{{\rm d}\beta_0U}\cr
&=&\frac{\alpha}{1-\alpha}
\frac{p_1^{\alpha-1}-p_0^{\alpha-1}}{p_0^\alpha+p_1^\alpha}
\,\frac{{\rm d}p_1}{{\rm d}\beta_0U}
\label{tresulttwolevel}
\end{eqnarray}
with
\begin{eqnarray}
\frac{{\rm d}p_1}{{\rm d}\beta_0U}
&=&\frac{1}{x_1}(1+(1-\alpha)p_1x_1)^{(\rho-\alpha)/(1-\alpha)}\cr
&=&\frac{1}{x_1}(\exp_\alpha(p_1x_1))^{\rho-\alpha}.
\end{eqnarray}

Let us shortly discuss these results.

The condition $T>0$ requires that $p_1<p_0$ holds.
The latter is also expected on physical grounds.
The equality $p_1=p_0$ is reached when temperature $T$ goes to infinity.
Expansion for small $p_1$, assuming $0<\alpha<1$, gives
\begin{equation}
\beta_0T\simeq \frac{1-\alpha}{\alpha}x_1p_1^{1-\alpha}
\end{equation}
This means that for small temperatures the population of the excited
level goes like $T^{1/(1-\alpha)}$. On the other hand,
if $\alpha>1$ then the limit $p_1=0$ is reached at a finite temperature $T_g$
given by
\begin{equation}
\beta_0T_g=(1-\frac{1}{\alpha})x_1.
\end{equation}
Below $T_g$ is $p_1=0$. Due to the cutoff in the probability
distribution the first excited level is not populated at all. 
See Figure 1.

In the special case $a=1$ one obtains directly from (\ref{2levx},\ref{2levp})
that $p_1=p_0\exp_\rho(\beta_0E_1)$.
The corresponding temperature is calculated using (\ref{spectemp}) and satisfies
\begin{equation}
\beta_0T=\frac{1}{\alpha}(1+(1-\rho)\beta_0U)
\end{equation}
In particular, as noted earlier, if $\rho=1$, then the special temperature
satisfies $\alpha\beta_0T=1$.
Hence, as a funtion of $\alpha T$, curves plotting $p_1$ as a function
of $T$ for different $\alpha$, but $\rho=1$, all intersect in the same point
--- see Figure 1.

\begin{figure}
\centerline{\psfig{figure=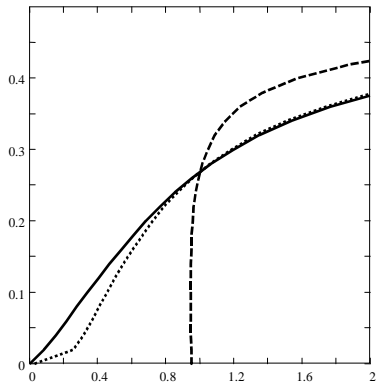,height=5cm}}
\centerline{\parbox[b]{5cm}}
{\caption
{Occupation of the excited level
as a function of  $\alpha T$, for $\rho=1$, $\beta_0E_1=1$, and $\alpha=0.5$
(solid line), resp.~$1.0$ (dotted), and $2.0$ (dashed). 
}
\label{fig:1}
}
\end{figure}

\subsection{Why nonlinear averages?}
\label{wna_sect}

We now show how nonlinear averages arise in a natural way.

Note that
\begin{eqnarray}
\phi_\alpha(\beta_0(E-U)/N)
&=&\frac{e^{(1-\alpha)\beta_0(E_\sigma-U)/N}-1}{1-\alpha}\cr
&=&\frac{e^{(1-\alpha)\beta_0E/N}-e^{(1-\alpha)\beta_0U/N}}
{(1-\alpha)e^{(1-\alpha)\beta_0U/N}}\cr
&=&\frac{1}{N}\,
\frac{\phi_{\alpha_N}(\beta_0E)-\phi_{\alpha_N}(\beta_0U)}{\phi'_{\alpha_N}(\beta_0U)}
\end{eqnarray}
with $1-\alpha_N=(1-\alpha)/N$.
In particular, the equation $\phi_\alpha(\beta_0(E-U)/N)\approx\epsilon$
appearing in the l.h.s.~of (\ref{largedev2}), can be rewritten as
\begin{equation}
\phi_{\alpha_N}(\beta_0E)-\phi_{\alpha_N}(\beta_0U)
\approx\phi'_{\alpha_N}(\beta_0U)\,\epsilon.
\end{equation}
This implies that (\ref{largedev2}) can be written as
\begin{eqnarray}
\hbox{Prob}\left((\phi_{\alpha_N}(\beta_0E)-\phi_{\alpha_N}(\beta_0U))/N\approx \epsilon
\right)
\sim \exp(-Nr_*(\epsilon))
\label{largedevext}
\end{eqnarray}
with
\begin{equation}
r_*(\epsilon)=r\circ\phi_\alpha^{-1}\left(\epsilon/\phi'_\alpha(\beta_0U/N)\right)
\end{equation}

The problem is now to find a probability
distribution $p_\sigma$ for which (\ref{largedevext}) is satisfied.
A candidate solution to this problem is obtained by maximizing entropy
with the constraint that
\begin{equation}
\sum_\sigma p_\sigma \phi_{\alpha_N}(\beta_0E_\sigma)=\phi_{\alpha_N}(\beta_0U)
\end{equation}
The latter condition can be written as
\begin{equation}
\langle \beta_0E\rangle =\beta_0U
\end{equation}
with average $\langle\cdot\rangle$ calculated as a KN-mean w.r.t.~the function
$\phi_{\alpha_N}$. This justifies the use of nonlinear KN-averages.

Remains the question which entropy should be optimized. We believe that
in many applications the right choice is either Shannon's
entropy or an $\alpha$-entropy of R\'enyi (or an equivalent entropy
like that of Tsallis). The reason for this belief is that these are the
only entropies with nice additivity properties.

\subsection{R\'enyi thermostatistics of the Curie-Weiss model}
\label{cw2_sect}

We reconsider the Curie-Weiss model, now with the function $\phi$ 
equal to $\phi_{\alpha,1}\equiv \phi_{\alpha}$. It is not the 
intention here to treat this model in detail, with arbitrary 
values of $\alpha$ and $\rho$. We rather want to to discuss the 
thermodynamic limit in the extensive $\rho=1$-case.

Introduce the notations $m=M/N$ and $u=U/N$, and
\begin{equation}
\gamma(m)=\exp((1-\alpha)\beta_0(Jm^2/2+hm+u)).
\end{equation}
The equilibrium probabilities are, assuming $0<\alpha<1$,
(see (\ref{ftdistr}) and the discussion of the previous subsection)
\begin{eqnarray}
p(M)
&=&{N\choose \frac{N+M}{2}}\frac{1}{Z}
\left[
\alpha_N+(1-\alpha_N)
\frac{\phi_{\alpha_N}(\beta_0E_\sigma)-\phi_{\alpha_N}(\beta_0U)}
{\phi'_{\alpha_N}(\beta_0U)\beta_0T}
\right]^{1/(\alpha_N-1)}\cr
&=&{N\choose \frac{N+M}{2}}\frac{1}{Z}
\left[\alpha_N -\frac{1}{\beta_0 T}\left(
1-\frac{1}{\gamma(m)}\right)
\right]^{-N/(1-\alpha)}\cr
&\sim&
\exp\bigg[-Ns(m)
\bigg],
\end{eqnarray}
with
\begin{eqnarray}
s(m)&=&s_\infty(m)+\frac{1}{1-\alpha}\ln\left[
1-\frac{1}{\beta_0T}\left(1-\frac{1}{\gamma(m)}\right)
\right]
\end{eqnarray}
(assuming $T>J$).
The first derivative of $s(m)$ is
\begin{eqnarray}
I'(m)&=&\frac{1}{2}\ln\frac{1+m}{1-m}
-\frac{\beta_0(Jm+h)}{1+(\beta_0T-1)\gamma(m)}
\end{eqnarray}
Let $m_*$ be the value of $m$ for which $s(m)$ is
maximal. It is a solution of $I'(m)=0$. Hence it satisfies
\begin{eqnarray}
m_*=\tanh\left(\frac{\beta_0(Jm_*+h)}{1+(\beta_0T-1)\gamma(m_*)}
\right)
\end{eqnarray}
But $\gamma(m_*)=1+\hbox{O}(1/N)$ holds for large $N$, so that
the previous equation simplifies to
the standard mean-field equation (\ref{mfeq}).

In case $T>J$ the rate function $s(m)$ does not differ
significantly from the standard result $s_\infty(m)-Jm^2/2T$.
If $T<J$ the difference consists in the first place of a constant
term. All together, the present treatment of the Curie-Weiss model
seems to yield results which are very similar to the standard one.

\section{Discussion}

We have presented a formalism of thermostatistics based on 
Kolmogorov-Nagumo averages and Hartley's information measure.
It treats the problem of maximizing average information $\langle I\rangle$
under the constraint that average energy $\langle \beta_0E\rangle$
has some given value $\beta_0U$. The KN-average $\langle\cdot\rangle$
is determined by a monotonic function $\phi$ and by probabilities $p_k$.
The same KN-average is used to define average energy and
to define entropy as an average information content.
In the present paper the function $\phi$ is kept constant
while the probabilities $p_k$ are varied. It could be interesting
to consider cases where $\phi$ is varied as well.

Of particular interest is the case when $\phi$ is an exponential 
function because then the average information $\langle I\rangle$ 
coincides with R\'enyi's entropy. As proved by R\'enyi, his 
$\alpha$-entropies, together with that of Shannon, are the only 
additive ones. Instead of exponential $\phi$ we use a slightly 
modified function (\ref{phirenyi}), for which $\langle I\rangle$ 
is still R\'enyi's entropy. We call the statistical formalism 
obtained in this way R\'enyi thermostatistics. It generalizes 
the standard formalism of Boltzmann-Gibbs. Note that R\'enyi 
thermostatistics is extensive in the sense that, when a system 
is composed of independent subsystems and the probabilities are 
of the product form, then both average information and average 
energy of the total system are the sum of the corresponding 
subsystem averages. From the examples of the Curie-Weis model 
(section \ref{cw2_sect}) and the two-level model (section 
\ref{2lev_sect}) it seems that R\'enyi thermostatistics behaves 
quite similar to the standard formalism. The most striking 
difference is the energy cut-off which can occur in case 
$\alpha>1$. When this happens then the probability of occupation 
of high-energy levels is exactly zero. In particular, this kind 
of equilibrium state is not a Gibbs measure.

We have tried to justify the use of KN-averages in statistical 
mechanics. In essence, a KN-average consists of a change of 
scale by means of a function $\phi$, followed by a statistical 
average using probabilities $p_k$, and a final back 
transformation using the inverse function $\phi^{-1}$. A basic 
ingredient of statistical mechanics is the equivalence of 
ensembles. A possible reason for combining the canonical 
ensemble with a change of scale is that in this way the 
equivalence of ensembles might be restored. We did not consider 
equivalence of ensembles directly, but argued on the level of 
fluctuations of energy. These should be exponentially small in 
order to apply the theory of large deviations. It is intuitively 
clear that a change of scale influences the scale of 
fluctuations. This is indeed confirmed by calculations on the 
Curie-Weiss model which we used as an example. Now, if the 
function $\phi$ is the function involved in the definition of 
R\'enyi's entropy, then scaled energy fluctuations can be 
expressed as fluctuations of the scaled energy. In this way we
can justify that scaled energy is the relevant quantity
of the canonical ensemble.

In a natural way Tsallis' entropy appears as a tool for 
calculating equilibrium averages in R\'enyi thermostatistics. 
This offers the opportunity to reuse knowledge from Tsallis-like 
thermostatistics. It shines also new light onto Tsallis' 
nonextensive thermostatistics because there is a one-to-one 
translation of problems from the one formalism to the other. 
There are many applications of the Tsallis formalism --- see 
e.g.~the review paper \cite{TC99}. An open question is how many 
of these applications have a natural formulation in terms of 
R\'enyi thermostatistics. As a first guess this would be the 
case whenever extensivity is important.

We consider also a further generalization of extensive R\'enyi 
thermostatistics to a non-extensive formalism. This extended 
formalism includes, besides the extensive case, also 
nonextensive Tsallis thermostatistics. In particular, an
argument has been presented which settles the problem of 
defining thermodynamic temperature in a nonextensive context.

A short version of the present paper has appeared in \cite{NC02}.

\begin{acknowledgement}
One of the authors (MC) wishes to thank the NATO for a research fellowship
enabling his stay at the Universiteit Antwerpen.
\end{acknowledgement}

\section*{Appendix}

Definition (\ref{defln}) of the $\alpha$-deformed logarithm $\ln_\alpha(x)$,
without the condition $x\ge 1$,
is what is found in literature \cite {TC94,BEP98}. The inverse function
$\exp_\alpha(x)$ is not defined for all $x$,
while we need it for all $x$ in section \ref{renyisect}.
Therefore we propose the following extended definition
\begin{eqnarray}
\ln_\alpha(x)&=&\frac{x^{1-\alpha}-1}{1-\alpha}
\qquad\hbox{ if }x\ge 1\cr
&=&\frac{x^{\alpha-1}-1}{\alpha-1}
\qquad\hbox{ if }0<x\le 1
\end{eqnarray}
(we assume $\alpha>0$, $\alpha\not=1$).
It satisfies $\ln_\alpha(1)=0$.
The deformed logarithm $\ln_\alpha(x)$
converges in the limit $\alpha=1$ to the natural logarithm
$\ln(x)$.

The derivative of $\ln_\alpha(x)$ is a continuous positive
function given by
\begin{eqnarray}
\frac{{\rm d}\,}{{\rm d}x}\ln_\alpha(x)
&=&x^{-\alpha}\qquad\hbox{ if }x\ge 1\cr
&=&x^{\alpha-2}\qquad\hbox{ if }0<x\le 1
\end{eqnarray}
In particular, this implies that $\ln_\alpha(x)$ is a strictly increasing
function of $x$.
Note that
\begin{equation}
-\frac{1}{\alpha - 1}<\ln_\alpha(x)<\frac{1}{\alpha - 1}
\qquad \hbox{ if }\alpha>1
\end{equation}
while for $0<\alpha<1$ the deformed logarithm takes on any real value.
Hence, the inverse function $\exp_\alpha(x)$ has a limited domain of
definition in the former case. One finds
\begin{eqnarray}
\exp_\alpha(x)
&=&\left[1+(1-\alpha)x\right]_+^{1/(1-\alpha)}
\qquad\hbox{ if }x\ge 0\cr
&=&\left[1+(\alpha-1)x\right]_+^{1/(\alpha-1)}
\qquad\hbox{ if }x\le 0
\end{eqnarray}
Clearly, $\exp_\alpha(0)=1$ and $\exp_\alpha(x)$ cannot be negative.
In the limit $\alpha=1$ the deformed exponential function
$\exp_\alpha(x)$ converges to the usual exponential function $\exp(x)$.

The derivative of $\exp_\alpha(x)$
\begin{eqnarray}
\frac{{\rm d}\,}{{\rm d}x}\exp_\alpha(x)
&=&\left(\exp_\alpha(x)\right)^\alpha
\qquad\hbox{ if }x\ge 0\cr
&=&\left(\exp_\alpha(x)\right)^{2-\alpha}
\qquad\hbox{ if }x\le 0
\end{eqnarray}
is a positive continuous function. In particular, this implies that
$\exp_\alpha(x)$ is a strictly increasing function on its domain
of definiton.

Finally, note that
\begin{equation}
\exp_\alpha(-x)=1/\exp_\alpha(x)
\qquad\hbox{ and }\quad
\ln_\alpha(1/x)=-\ln_\alpha(x).
\label{expneg}
\end{equation}


\end{document}